# A simple self-interaction correction to RPA-like correlation energies

Tim Gould[1], Adrienn Ruzsinszky[2*], and John P. Perdew[2]

[1]*Qld Micro- and Nanotechnology Centre,*

*Griffith University, Nathan, Qld 4111, Australia*

[2] *Department of Physics, Temple University, Philadelphia, PA, 19122, USA*

## Abstract

The random phase approximation (RPA) is exact for the exchange energy of a many-electron ground state, but RPA makes the correlation energy too negative by about 0.5 eV/electron. That large short-range error, which tends to cancel out of iso-electronic energy differences, is largely corrected by an exchange-correlation kernel, or (as in RPA+) by an additive local or semilocal correction. RPA+ is by construction exact for the homogeneous electron gas, and it is also accurate for the jellium surface. RPA+ often gives realistic total energies for atoms or solids in which spin-polarization corrections are absent or small. RPA and RPA+ also yield realistic singlet binding energy curves for $H_2$ and $N_2$, and thus RPA+ yields correct total energies even for spin-unpolarized atoms with fractional spins and strong correlation, as in stretched $H_2$ or $N_2$. However, RPA and RPA+ can be very wrong for spin-polarized one-electron systems (especially for stretched $H_2^+$), and also for the spin-polarization energies of atoms. The spin-polarization energy is often a small part of the total energy of an atom, but important for ionization energies, electron affinities, and the atomization energies of molecules. Here we propose a computationally efficient *generalized* RPA+ (gRPA+) that changes RPA+ only for spin-polarized systems by making gRPA+ exact for all one-electron densities, in the same simple semilocal way that the correlation energy densities of many meta-generalized gradient approximations are made self-correlation free. By construction, gRPA+ does not degrade the exact RPA+ description of jellium. gRPA+ is found to greatly improve upon RPA and RPA+ for the ionization energies and electron affinities of light atoms. Many versions of RPA with an approximate exchange-correlation kernel fail to be exact for all one-electron densities, and they can also be self-interaction corrected in this way.

*Corresponding author

## I. Introduction

For correlation energies, the random-phase approximation (RPA)[1,2] has attracted significant attention over the past decade or two due to its ability to treat different types of interactions (e.g., dispersion, metallic, covalent and ionic interactions) seamlessly, at a cost that is tractable for moderately large systems. For this reason, RPA has been used to provide "silver standard" benchmarking for electronic systems where more refined (and thus expensive) approaches are intractable[3–7].

RPA is typically used as a post-density-functional-theory (DFT) method for calculating energies. That is, one first performs an approximate DFT calculation, e.g., using the Perdew-Burke-Ernzerhof[8] (PBE) functional, and then uses the resulting orbitals and their energies to calculate the non-interacting (Kohn-Sham) response function $\chi_0$, and from it the RPA correlation energy

$$E_{cRPA} = \int_0^1 d\lambda \int \frac{drdr'}{2|r-r'|} n_{2c,\lambda}^{RPA}(r,r') \ , \qquad (1)$$

where $n_{2c,\lambda}^{RPA} = \int_0^\infty \frac{d\omega}{\pi} [\chi_\lambda - \chi_0]$ is the contribution to the pair-density from Coulomb correlation. The random-phase approximation involves approximating the interacting response function as $\chi_\lambda = \chi_0 + \lambda \chi_0 * U * \chi_\lambda$. Stars indicate convolution over interior variables and $U = 1/|r-r'|$ is the Coulomb potential. We can also rewrite (1) as an integral

$$E_{cRPA} = \int dr n(r) \epsilon_{cRPA}(r) \qquad (2)$$

over a correlation energy density. Here $n = n_\uparrow + n_\downarrow$ is the electron density. The correlation energy per electron at $r$ is

$$\epsilon_{cRPA}(r) = \int \frac{dr'}{2|r-r'|} \int_0^1 d\lambda \frac{n_{2,\lambda}^{RPA}(r,r')}{n(r)} \ . \qquad (3)$$

RPA typically does a decent job of estimating energy differences between similar species, such as the interactions between two-dimensional layers[4]. RPA yields reasonable structural properties for non-magnetic solids[9] and captures the static correlation in the dissociating $H_2$ [10,11]. It does a much poorer job of treating absolute energies, or of predicting energies where the fundamental properties of the different species are different [6,7,12,13]. For example, while RPA is relatively exact for the correlation energy in the high-density limit of the uniform gas, it makes the correlation energy per electron too negative by about 0.5 eV or 11.5 kcal/mol[1,2]. A similar result holds for atoms[14]. Furthermore the failure of RPA for dissociating $H_2^+$ is dramatic[15]. Perhaps most disappointing has been the performance[9] of RPA for the atomization energies of molecules, which is not better than that of the much less expensive PBE generalized gradient approximation[8].

Significant work has been dedicated to improving energies by modelling the exchange and correlation kernel $f_{xc}^\lambda(r, r'; \omega)$ that, when added to the Coulomb interaction $\lambda U(r' - r)$, would in principle make RPA exact via the introduction of a properly-screened or damped electron-electron interaction (e.g., refs. 16–21). Besides the standard particle-hole RPA, there is also a self-correlation-free particle-particle RPA[22]. Such "beyond-RPA" methods often offer some improvement over RPA, but have the unfortunate property of typically being much more computationally demanding than RPA.

An alternate and low-cost route to overcome this shortcoming was provided by Kurth and Perdew[23]. They recognised that the main source of error in the RPA energy was from short-ranged correlations, which could be corrected locally. They thus proposed that the overall correlation energy could be improved by setting

$$E_{cRPA+} = \int dr\, n(r)\, \epsilon_{cRPA+}(r), \qquad (4)$$

$$\epsilon_{cRPA+}(r) = \epsilon_{cRPA}(r) + \epsilon_{cLSDA}(n_\uparrow(r), n_\downarrow(r)) - \epsilon_{cLRPA}(n_\uparrow(r), n_\downarrow(r)), \qquad (5)$$

to provide a local correction to the RPA energy based on its error for the homogeneous electron gas. Here, the local spin density approximation (LSDA) and LRPA energies are taken from the homogeneous electron gas using respectively exact and RPA correlation energies per electron, as parametrised in ref. 24.

Using (4) yields more accurate correlation energies for many systems. Unfortunately, it does not offer significant improvements to reaction energies when the reactants differ significantly from the products, especially when the effective number of electrons changes. This lack of success reflects non-systematic errors in how the semi-local correction improves the properties of fundamentally different systems, such as neutrals and their ions.

In this work, we thus propose a straightforward modification to (4), that significantly improves its treatment of ionization energies and electron affinities, and improves its overall accuracy. This involves calculating a new self-interaction corrected generalized RPA+ (gRPA+) energy expression:

$$E_{cgRPA+} = \int d\mathbf{r}\, n(\mathbf{r}) g(\zeta(\mathbf{r}), \beta(\mathbf{r})) \epsilon_{cRPA+}(\mathbf{r}), \tag{6}$$

where $g(\zeta,\beta)$ is a meta-GGA[25]-like damping factor that zeroes out the correlation energy density in one-electron regions. $g(\zeta,\beta)$ depends upon the spin-polarization function $\zeta = \frac{n_\uparrow - n_\downarrow}{n}$, and the switching variable

$$0 \leq \beta = \frac{\tau - \tau_W}{\tau + \tau_{HEG}} \leq 1 \tag{7}$$

that indicates when a system is one- or two-electron like ($\beta \to 0$), homogeneous electron gas or HEG-like ($\beta \approx 1/2$), or involves overlapping shells ($\beta \to 1$)[26]. Here, $\tau = \frac{1}{2}\sum_i f_i |\nabla \phi_i|^2$ is the positive orbital kinetic energy density that depends on the occupied Kohn-Sham orbitals $\phi_i$; $\tau_W = |\nabla n|^2/8n$ is the von Weizsaecker approximation, which is exact for one- and two-electron densities; and $\tau_{HEG} = (3/10)(3\pi^2)^{2/3} n^{5/3}$ is the Thomas-Fermi kinetic energy approximation, which is exact for uniform densities.

In the next section we propose a model for the factor $g$ in Eq. (4), and justify it on physical and practical grounds. Then, we apply it to various tests.

## II. MODEL FOR THE FACTOR $g(\zeta, \beta)$ IN EQ. (6); IONIZATION POTENTIALS OF ATOMS

gRPA+ seeks to improve RPA+ by modifying the RPA+ correlation energy density locally. Before presenting our model for the damping factor $g(\zeta, \beta)$, we summarize what is known:

1. RPA+ works well when a system is spin-unpolarized ($\zeta = 0$), where we expect $g = 1$;
2. The correlation energy density should vanish in one-electron regions (with $\beta \approx 0$ and $|\zeta| \approx 1$), where we expect $g = 0$
3. RPA+ is exact when the system is HEG-like (with $\beta \approx 1/2$), and thus we expect $g = 1$ in HEG-like regions;
4. Intermediate-range van der Waals interactions[25] occur in density-overlap regions where $\beta$ is close to 1, so we expect $g = 1$ there.

A simple model that accommodates these various physical conditions is

$$g(\zeta,\beta) = 1 - \zeta^2 h(\beta), \tag{8}$$

provided $h(\beta = 0) = 1$ and $h(1/2) = 0$. Smoothness of $h$ is another desirable property, both for computational and physical reasons. One such model $h$ that meets the desired criteria is

$$h(\beta) = \begin{cases} \exp\left(-\dfrac{c\beta}{\frac{1}{2}-\beta}\right), & \beta \leq \frac{1}{2} \\ 0 & \beta > \frac{1}{2} \end{cases} \quad (9)$$

Here $c > 0$ is a parameter, and $c \to \infty$ recovers RPA+. For non-zero $\zeta$, this model gives maximum modification of RPA+ at $\beta = 0$, and no modification for HEG-like regions or in regions where shells overlap.

With the model chosen, we must select a value of $c$ that makes it work effectively. Our goal is to remove non-systematic errors. So, to choose $c$, we seek to minimize the mean absolute error (MAE) of the correlation contribution to the ionization potentials of various atoms, specifically Be, C, N, O, F and Na. That is, we define

$$MAE_{\text{IP}} = \frac{1}{6}\Sigma_{Z\in\{4,6,7,8,9,11\}}\left|IP_c^{\text{gRPA+}}(Z) - IP_c^{\text{exact}}(Z)\right| \quad (10)$$

for valid values of $c$, and seek to find its optimal value $c_{\text{opt}} = \text{argmin}_c\, MAE_{\text{IP}}(c)$. Here $IP_c(Z) = E_c(Z,Z\text{-}1) - E_c(Z,Z)$ where $E_c^{\text{exact}}(Z,N)$ is the exact correlation energy of $N$ electrons in a nuclear potential $-Z/r$, and $E_c^{\text{gRPA+}}(Z,N)$ is its equivalent calculated using Eq. (8) in Eq. (6). Exact values for $IP_c$ are taken from Chakravorty et al[27].

Our calculations are carried out in the atomic code pyAtom, which is a python/scipy/numpy implementation of DFT in spherical geometries. We construct the Kohn-Sham orbitals and energies of the atoms and atomic ions using a spherically-symmetric Kohn-Sham potential produced by ensemble averaging[14,28], that reproduces by construction correct density variables (e.g., $\zeta, \beta$) and other properties (e.g, $\chi_0, \varepsilon_{cRPA}$) of both closed-shell and open-subshell cases. This approach has previously been shown to reproduce well the properties of atoms[14,29]. The code is available on request.

Our tests indicate that $MAE_{\text{IP}}(c)$ attains its minimum 3.2 kcal/mol for $c_{\text{opt}} \to 0^+$. However, such a value is unphysical as it corresponds to a step function for $\beta < 1/2$. We thus choose instead the value $c_{\text{opt}} = 0.2$, which gives an error less than 0.5 kcal/mol above the absolute minimum, but which retains a physically reasonable form for $g$, as shown in Figure 1. It thus attains a good balance between physical constraints and error mitigation.

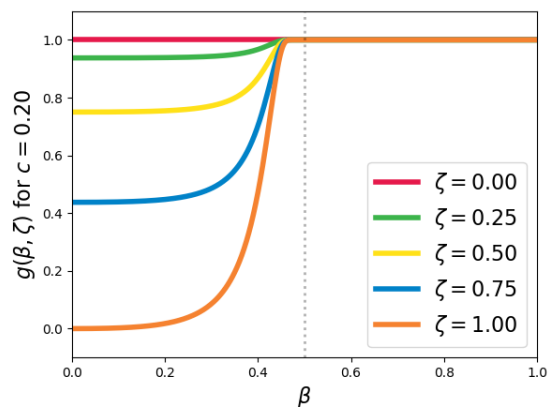

FIG. 1. The damping factor $g(\zeta, \beta)$ of Eqs. (6) and (8). Here, we selected $c_{opt}$ = 0.2. The vertical line shows $\beta$=0.5.

Using $c_{opt}$ = 0.2 gives MAE$_{IP}$(0.2) = 3.6 kcal/mol. For comparison, the equivalent error in RPA+ is 8.6 kcal/mol, more than twice as large and only a tiny improvement on RPA, with 9.2 kcal/mol, despite the fact that RPA+ improves absolute energies significantly (see later discussion). Thus, introducing the damping factor has the desired effect of significantly improving results for energy differences between unlike systems. Values for IP$_c$ for the systems used in the optimization, and their errors under the different methods, are reported in Table I.

TABLE I. Correlation contributions IP$_c$ to ionization potentials IP and their errors for RPA, RPA+ and gRPA+ with $c_{opt}$ = 0.2. All energies in kcal/mol. Mean absolute errors (MAE) are also reported.

| Element | IP$_c$ Exact | Error RPA | RPA+ | gRPA+ |
|---|---|---|---|---|
| Be | 29.4 | 8.2 | 17.4 | 7.8 |
| C | 11 | 8.1 | -3.8 | -5.3 |
| N | 13.6 | 15.3 | 1.5 | 1.5 |
| O | 40.2 | -0.9 | -10.6 | -6.2 |
| F | 40.2 | 9.1 | -2.7 | -0.9 |
| Na | 3.8 | 13.5 | -15.8 | 0.3 |
| MAE | | 9.2 | 8.6 | 3.6 |

## III. ELECTRON AFFINITIES OF ATOMS; CORRELATION ENERGIES OF ATOMS AND IONS

Table II tests gRPA+ for correlation contributions $EA_c(Z) = E_c(Z,Z) - E_c(Z, Z+1)$ to the electron affinities EA of the atoms studied in Table I. (See also ref. 29.) The results are qualitatively like those for the ionization potentials in Table I, suggesting that gRPA+ is not skewed to electron removal, but can also handle the generally more difficult case of electron addition.

TABLE II. Correlation contributions $EA_c$ to electron affinities EA and their errors for RPA, RPA+ and gRPA+ with $c_{opt} = 0.2$. All energies in kcal/mol. Mean absolute errors (MAE) are also reported.

|         | $EA_c$ |      | Error |       |
|---------|--------|------|-------|-------|
| Element | Exact  | RPA  | RPA+  | gRPA+ |
| C       | 16.6   | 16.1 | -1.6  | 2.6   |
| N       | 44.7   | -1.3 | -8.8  | -5.8  |
| O       | 45.9   | 8.1  | -2.8  | -1.1  |
| F       | 47.1   | 17.0 | 5.8   | 6.3   |
| Na      | 15.0   | 9.7  | 21.4  | 5.4   |
| MAE     |        | 10.4 | 8.1   | 4.2   |

We now take our optimized damping factor $g$ of Eqs. (6) and (8), and apply it to a larger set of atoms and ions to see how well it works. We expand our set of elements to include Mg, Al and P, and include double and triple cations as well as some anions in our tests. Results for the total correlation energy are shown in Figure 2, and values are reported in Table III.

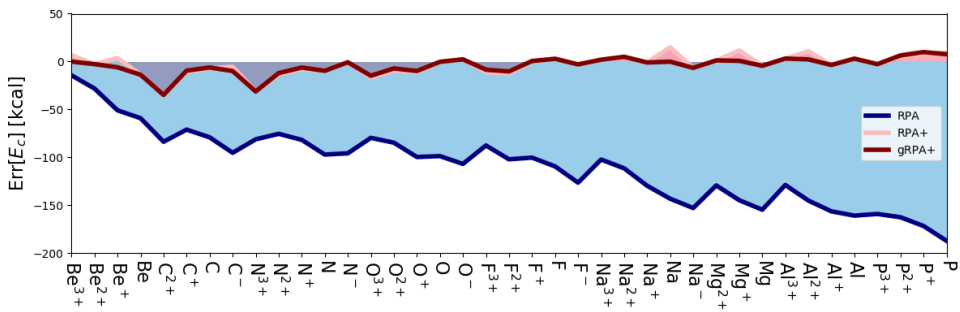

FIG. 2. Correlation energy errors (kcal/mol) for the full set of atoms and ions, using Eq. (8) in Eq. (6).

It is immediately clear from the figure that the greatest improvement to energies comes from RPA+. However, RPA+ correlation energies are fairly jagged across cations of the same species, which leads to non-systematic errors once energy differences are considered. gRPA+, while imperfect, tends to smooth out these errors and thus reduces the error in ionization potentials and electron affinities.

It is apparent from Table III that the gRPA+ approximation works best for atoms with open-shells. This is the case for carbon, nitrogen and sodium. For spin-unpolarized atoms, the RPA+ and gRPA+ approximations agree. For some spin-polarized atoms such as aluminium and sodium, the first ionization energy of gRPA+ is close to that of RPA+.

Interesting errors, which cannot be fixed by gRPA+, are those for the Be isoelectronic series: Be, $C^{2+}$ and $N^{3+}$ in our set. In all these cases RPA+ deviates from its usual ability to correct closed-shells well. We speculate that this might reflect that the near-degeneracy of $2s$ and $2p$ orbitals is inadequately captured in the uncorrected long-range part of RPA.

TABLE III. Correlation energy $E_c$ and errors for RPA, RPA+ and gRPA+ with $c_{opt}$ = 0.2. All energies in kcal/mol.

|   |   | $E_c$ | | Error | | |
|---|---|---|---|---|---|---|
| Z | N | Exact | gRPA+ | RPA | RPA+ | gRPA+ |
| 4 | 1 | 0.0 | 0.0 | -14.3 | 6.6 | 0.0 |
| 4 | 2 | -27.6 | -30.3 | -28.1 | -2.6 | -2.6 |
| 4 | 3 | -29.7 | -35.8 | -51.0 | 3.6 | -6.0 |
| 4 | 4 | -59.2 | -73.0 | -59.2 | -13.8 | -13.8 |
| 6 | 4 | -69.9 | -104.6 | -83.6 | -34.8 | -34.8 |
| 6 | 5 | -87.1 | -95.5 | -71.0 | -10.9 | -8.4 |
| 6 | 6 | -98.2 | -101.2 | -79.0 | -7.0 | -3.1 |
| 6 | 7 | -114.8 | -120.5 | -95.1 | -5.4 | -5.7 |
| 7 | 4 | -88.2 | -119.3 | -81.1 | -31.2 | -31.2 |
| 7 | 5 | -94.5 | -105.2 | -75.3 | -13.7 | -10.8 |
| 7 | 6 | -104.6 | -107.6 | -81.7 | -8.1 | -3.0 |
| 7 | 7 | -118.2 | -122.7 | -97.0 | -9.6 | -4.5 |
| 7 | 8 | -162.9 | -161.6 | -95.7 | -0.8 | 1.3 |
| 8 | 5 | -101.0 | -114.5 | -79.6 | -16.7 | -13.5 |
| 8 | 6 | -109.8 | -113.6 | -84.6 | -9.5 | -3.8 |
| 8 | 7 | -121.8 | -126.3 | -99.5 | -11.2 | -4.5 |
| 8 | 8 | -161.9 | -160.3 | -98.6 | -0.5 | 1.7 |
| 8 | 9 | -207.8 | -205.0 | -106.7 | 2.3 | 2.8 |
| 9 | 6 | -114.2 | -119.1 | -87.5 | -11.2 | -4.9 |
| 9 | 7 | -124.9 | -129.8 | -101.9 | -12.3 | -4.9 |
| 9 | 8 | -163.8 | -161.3 | -100.2 | 0.2 | 2.5 |
| 9 | 9 | -204.0 | -200.7 | -109.3 | 2.9 | 3.3 |
| 9 | 10 | -251.0 | -254.0 | -126.3 | -2.9 | -2.9 |
| 11 | 8 | -168.2 | -164.3 | -102.2 | 1.4 | 3.9 |
| 11 | 9 | -205.2 | -199.9 | -111.3 | 4.9 | 5.4 |
| 11 | 10 | -244.1 | -245.1 | -129.6 | -0.9 | -0.9 |
| 11 | 11 | -247.9 | -249.1 | -143.1 | 14.8 | -1.2 |
| 11 | 12 | -262.9 | -269.5 | -152.8 | -6.6 | -6.6 |
| 12 | 10 | -244.8 | -243.5 | -129.2 | 1.2 | 1.2 |
| 12 | 11 | -251.0 | -250.9 | -144.6 | 11.5 | 0.1 |
| 12 | 12 | -274.9 | -279.3 | -154.5 | -4.4 | -4.4 |
| 13 | 10 | -245.4 | -242.2 | -128.7 | 3.2 | 3.2 |
| 13 | 11 | -254.2 | -252.2 | -145.0 | 10.7 | 2.0 |
| 13 | 12 | -283.7 | -287.2 | -156.2 | -3.5 | -3.5 |
| 13 | 13 | -295.0 | -291.2 | -160.7 | 3.1 | 3.8 |
| 15 | 12 | -296.2 | -298.9 | -159.0 | -2.7 | -2.7 |
| 15 | 13 | -313.2 | -305.8 | -162.4 | 4.9 | 7.4 |
| 15 | 14 | -327.6 | -314.5 | -171.5 | 8.5 | 13.1 |
| 15 | 15 | -338.9 | -325.5 | -187.0 | 9.0 | 13.4 |
| MAE | | | | 108.1 | 8.5 | 6.2 |

## IV. CONCLUSIONS

Our results and previous ones suggest that RPA+ correlation energy densities are accurate for spin-unpolarized density regions ($|\zeta| = 0$) and for density regions in which various orbital shapes are strongly overlapped ($\beta > 0.4$), but not for density regions that are nearly one-electron-like. For the latter regions, we have proposed a self-interaction corrected or generalized RPA+ (gRPA+). We have shown that gRPA+ greatly improves the ionization potentials and electron affinities of atoms, in which spin-polarization effects play an important role. In future work, we plan to build this self-interaction correction into a molecular code to test it for atomization energies, where spin-polarization effects are also important. Table III shows that for most spin-polarized neutral atoms the gRPA+ correlation energies are less negative than the RPA+ values, and this goes in the right direction to reduce the RPA+ underbinding[12] of molecules.

In the RPA+ correlation energy density, the RPA term and its local short-range correction term need the same kinds of damping factors, but the $c$ parameters can differ from one term to the other.

The integrand of the RPA double integral is symmetric under interchange of **r** and **r'**, and, while our choice of Eq. (6) is equivalent to a symmetrized damping factor $[g(\mathbf{r}) + g(\mathbf{r'})]/2$, another possibility is $\sqrt{g(\mathbf{r})g(\mathbf{r'})}$. We will investigate these possibilities in future work.

The version of RPA+ we have employed here is one in which the correction to RPA is evaluated in the local spin-density approximation (LSDA). There is also a sophisticated generalized gradient approximation version[30] of RPA+. It is numerically close to the simpler LSDA, and will also be tested in future work.

The computational demands of RPA, RPA+, and gRPA+ were nearly the same in our work. We suspect that gRPA+ should be obtainable at a cost similar to that of resolution-of-the-identity based RPA implementations.

Exchange-correlation kernel corrections to RPA can be more accurate but more expensive than RPA+. They can also be self-interaction corrected via our general approach, probably with a larger parameter $c>0.2$.

## ACKNOWLEDGMENT

AR acknowledges National Science Foundation (NSF) support under Grant No. DMR-1553022. JPP acknowledges NSF support under Grant No. DMR-1607868.